%% file: main.tex
\documentclass[10pt,twocolumn,letterpaper]{article}

\usepackage{cvpr}              % To produce the CAMERA-READY version
\usepackage{booktabs}
\usepackage{tabularx}
\usepackage{amssymb}
\usepackage{pifont}
\usepackage[table]{xcolor}
\usepackage{subcaption}
\usepackage{amsmath}
\usepackage{adjustbox}
\usepackage{multirow}
\usepackage{array}
\newcommand{\cmark}{\ding{51}}
\newcommand{\xmark}{\ding{55}}
\input{preamble}
\definecolor{cvprblue}{rgb}{0.21,0.49,0.74}
\usepackage[pagebackref,breaklinks,colorlinks,allcolors=cvprblue]{hyperref}

\title{MoLT: Mixture of Layer-Wise Tokens for Efficient Audio-Visual Learning}

\author{
Kyeongha Rho\textsuperscript{*1}\quad\quad
Hyeongkeun Lee\textsuperscript{*1}\quad\quad
Jae Won Cho\textsuperscript{2}\quad\quad
Joon Son Chung\textsuperscript{1}
\\
\textsuperscript{1} KAIST\quad\quad\quad
\textsuperscript{2} Sejong University \\
\texttt{\{khrho325,lhk528,joonson\}@kaist.ac.kr}\quad
\texttt{chojw@sejong.ac.kr} \\
}

\begin{document}
\maketitle

\renewcommand{\thefootnote}{}
\footnotetext{\textsuperscript{*} equal contribution}

\input{sec/0_abstract}

\input{sec/1_intro}
\input{sec/2_rel_works}

\input{sec/3_method}

\input{sec/4_exp}

\input{sec/5_conclusion}

{
    \small
    \bibliographystyle{ieeenat_fullname}
    \bibliography{shortstrings,main}
}


\end{document}

%% file: sec/0_abstract.tex
\begin{abstract}

In this paper, we propose Mixture of Layer-Wise Tokens (MoLT), a parameter- and memory-efficient adaptation framework for audio-visual learning.
The key idea of MoLT is to replace conventional, computationally heavy sequential adaptation at every transformer layer with a parallel, lightweight scheme that extracts and fuses layer-wise tokens only from the late layers.
We adopt two types of adapters to distill modality-specific information and cross-modal interaction into compact latent tokens in a layer-wise manner.
A token fusion module then dynamically fuses these layer-wise tokens by taking into account their relative significance.
To prevent the redundancy of latent tokens, we apply an orthogonality regularization between latent tokens during training.
Through the systematic analysis of the position of adaptation in the pre-trained transformers, we extract latent tokens only from the late layers of the transformers.
This strategic adaptation approach avoids error propagation from the volatile early-layer features, thereby maximizing the adaptation performance while maintaining parameter and memory efficiency.
Through extensive experiments, we demonstrate that MoLT outperforms existing methods on diverse audio-visual benchmarks, including Audio-Visual Question Answering, Audio-Visual Segmentation, and Audio-Visual Event Localization. 

\end{abstract}

%% file: sec/1_intro.tex
\section{Introduction}
\label{sec:intro}

\input{figs/fig1_teaser}

Audio-visual learning has garnered significant attention in recent years, driven by the goal of enabling machines to emulate human intelligence in processing and analyzing complex, simultaneous audio and visual signals from real-world scenarios. 
% This field aims to comprehensively perceive and understand the world by leveraging the complementary nature of auditory and visual cues. 
The importance of audio-visual learning has been demonstrated across various downstream tasks, including Audio-Visual Question Answering (AVQA)~\cite{yang2022avqa,li2023progressive,park2024enhancing,kim2025qatiger}, Audio-Visual Segmentation (AVS)~\cite{zhou2022avs,li2023catr,gao2024avsegformer,liu2024frameexploit,chen2024unravel}, and Audio-Visual Event localization (AVE)~\cite{tian2018audio,xia2022crossmodal,wu2019dual,zhou2025towards}.
While the advent of large-scale pre-trained models~\cite{liu2022swin,chen2022htsat} has accelerated progress, it has also introduced a significant challenge that fully fine-tuning the entire parameter set of these large-scale models for each specific downstream task demands prohibitive computational costs, making it impractical for most applications.

To address this computational barrier, Parameter-Efficient Fine-Tuning (PEFT) methodologies~\cite{lin2023vision,duan2023cross,cheng2024avmoe} offer a promising alternative by freezing the parameters of a pre-trained model and introducing only a small number of new, trainable parameters.
However, previous audio-visual PEFT methods do not take into account memory efficiency, leading to substantially increased memory overhead and training time.
This memory inefficiency prevents training the model with very limited computational resources.
Recent work~\cite{zhou2025mettle} proposes to alleviate this issue by distilling each transformer layer into a set of learnable latent tokens in parallel.
However, it focuses only on modality-specific information, overlooking cross-modal interaction.
Furthermore, it uses a simple average of the extracted latent tokens for the final prediction, ignoring their distinct roles and relative importance.
These shortcomings lead to degraded adaptation performance on complex reasoning tasks, including AVQA.

In this paper, we explore the practical sweetspot among the three main axes of efficient audio-visual learning: parameter efficiency, memory efficiency, and generalizability of adaptation across the downstream tasks.
To this end, we propose \textbf{MoLT}, a novel parameter- and memory-efficient audio-visual learning framework that intelligently fuses a ``\textbf{M}ixture \textbf{o}f \textbf{L}ayer-Wise \textbf{T}okens". 
MoLT is motivated by two key insights: (1) modality-specific information and cross-modal interaction play a complementary role, necessitating an approach that effectively utilizes them in a balanced manner, and (2) representations in transformers become increasingly abstract with depth, with later layers capturing high-level semantics.
Specifically, for frozen audio and visual backbones, we introduce the Feature Distillation Module (FDM), a mixture of two adapter types: uni-modal and cross-modal distillation adapters.
These adapters extract compact latent tokens that summarize modality-specific information and cross-modal interaction from each transformer layer, respectively.
Then, a router assigns weights to these adapters, creating a final latent token for each layer by computing a weighted sum of their respective outputs.
Furthermore, we introduce an MLP-based Token Fusion Module (TFM) to effectively fuse the layer-wise latent tokens.
It estimates the relative importance of each latent token, and importance weights are then used to compute a final weighted sum of all layer-wise tokens for the final prediction.
To prevent trivial solutions where the latent tokens extracted from each layer become redundant and excessively similar, we apply an orthogonality regularization between latent tokens during training.

Crucially, we extract latent tokens only from the late layers of transformer backbones, unlike the previous approaches that adapt every transformer layer. 
Representations change drastically in the early transformer layers; if adapters in the early layers distort them, the error propagates and degrades the adaptation performance.
By concentrating adaptation on the late transformer layers, MoLT prevents early-stage error accumulation and improves adaptation performance while reducing both parameters and memory overhead.
To the best of our knowledge, we provide the first systematic analysis that pinpoints which layers of pre-trained transformers benefit most from adaptation.

Through extensive experiments, we demonstrate that the proposed approach outperforms existing methods on various audio-visual downstream tasks, including Audio-Visual Question Answering (AVQA), Audio-Visual Segmentation (AVS), and Audio-Visual Event Localization (AVE).

Our contributions are summarized as follows:
\begin{itemize}
% \item We propose MoLT, a novel parameter- and memory-efficient audio-visual learning framework that intelligently fuses a ``Mixture of Layer-Wise Tokens". 
\item We propose \textit{MoLT}, a novel framework for efficient audio-visual learning that distills each transformer layer into compact latent tokens via a mixture of uni-modal and cross-modal adapters, balancing modality-specific cues and cross-modal interaction.

\item We introduce the Token Fusion Module that effectively fuses the layer-wise latent tokens by considering their relative importance. 
Furthermore, Token Orthogonality Regularization prevents the latent tokens from becoming overly redundant, promoting distillation of complementary information across the latent tokens.

\item We provide a systematic analysis to demonstrate that our late-layer adaptation strategy is more effective than the conventional all-layer adaptation approach, while ensuring both parameter and memory efficiency.

\item MoLT achieves the state-of-the-art performance on diverse audio-visual downstream tasks, including Audio-Visual Question Answering, Audio-Visual Segmentation, and Audio-Visual Event Localization. 

\end{itemize}

%% file: figs/fig1_teaser.tex
\begin{figure}[t!]
    \centering
    \includegraphics[width=0.9\linewidth]{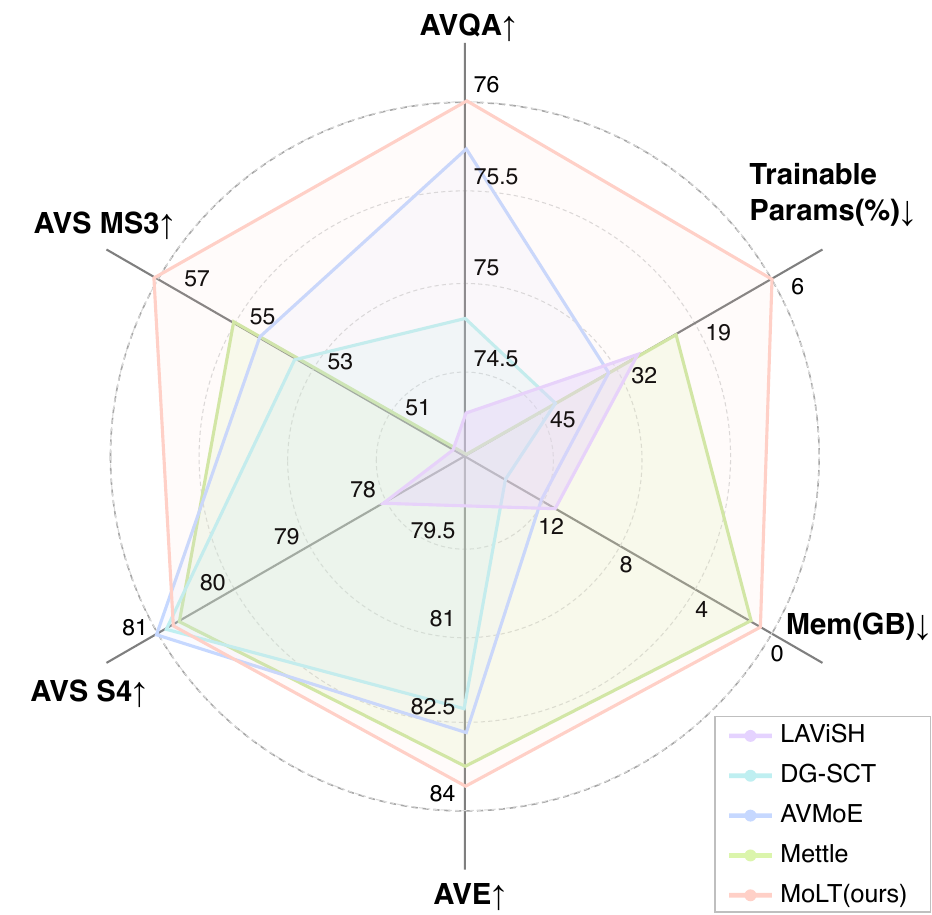}
    \caption{The proposed MoLT achieves the state-of-the-art performance on various audio-visual downstream tasks, while ensuring efficiency in terms of both parameter and memory.}
    \label{fig:teaser}
    \vspace{-1.2em}
\end{figure}

%% file: sec/2_rel_works.tex
\section{{Related Work}} 
\label{sec:formatting}

\subsection{{Audio-Visual Learning}}
Audio-visual learning aims to achieve a holistic understanding of multimodal scenes by utilizing both audio and visual information~\cite{yang2022avqa,li2023progressive,park2024enhancing,kim2025qatiger,zhou2022avs,li2023catr,gao2024avsegformer,liu2024frameexploit,chen2024unravel,tian2018audio,xia2022crossmodal,wu2019dual,zhou2025towards}.
In particular, this paper investigates three core audio-visual tasks: Audio-Visual Event localization (AVE), Audio-Visual Segmentation (AVS), and Audio-Visual Question Answering (AVQA). AVE~\cite{tian2018audio} aims to classify event categories and localize temporal segments where the corresponding audio and visual events occur, and previous works~\cite{wu2019dual, xu2020cross, xia2022crossmodal, zhou2025towards} employ cross-modal modeling to effectively leverage multimodal cues obtained from modality-specific pre-trained models. In addition to localizing events, AVS~\cite{zhou2022avs} focuses on segmenting sounding objects by producing pixel-level masks that visually ground the audio source. The segmentation in prior studies~\cite{li2023catr, gao2024avsegformer, liu2024frameexploit, ma2024steppingstones, chen2024unravel} is commonly guided by learning audio-visual correspondences. AVQA~\cite{yang2022avqa} addresses the task of answering various types of questions audio-visual content through a comprehensive understanding of both modalities. Subsequent works~\cite{li2023progressive, chen2024globallocal, park2024enhancing} focus on enhancing the model’s reasoning capability to improve answer accuracy.

Most works in audio-visual learning apply large-scale audio-visual pre-trained models to downstream tasks. Despite their strong performance, these models do not fully leverage complementary information across modalities and incur high computational and memory costs. In this work, we introduce a parameter- and memory-efficient audio-visual learning framework that addresses these limitations.

\input{figs/fig2_main}

\subsection{{Parameter-Efficient Fine-Tuning}}
Parameter-Efficient Fine-Tuning (PEFT) has emerged as a promising strategy for adapting pre-trained models to downstream tasks, as it updates only a small fraction of training parameters, thereby avoiding full-model fine-tuning and significantly reducing both training and memory overhead.
With extensive development in the natural language processing domain~\cite{houlsby2019parameter, bapna2019simple, karimi2021compacter, hu2022lora, lester2021power, gu2022ppt, li2021prefix}, PEFT has expanded to a wide range of modalities, including computer vision~\cite{gao2023compositional,park2024fair,yang2023fine,wang2025learning} and vision-language tasks~\cite{yao2024tcp,zhang2024dept,khattak2023self,roy2024copromt,zhou2024label,khattak2023maple,huang2023vop,chen2023plot}.

In the audio-visual domain, LAVisH~\cite{lin2023vision} inserts cross-modal adapters into every transformer layer of the visual and audio backbones to enable efficient cross-modal fusion with minimal trainable parameters. Building on adapter-based PEFT, DG-SCT~\cite{duan2023cross} enhances audio-visual representation learning through a dual-guided spatial–channel–temporal attention mechanism. AVMoE~\cite{cheng2024avmoe} extends this line of work with a layer-wise mixture-of-experts architecture that integrates uni-modal and cross-modal adapters, and dynamically routes the optimal expert combination at each layer. 
Beyond adapter designs, Mettle~\cite{zhou2025mettle} alleviates memory inefficiency in audio-visual PEFT by distilling compact audio and visual meta-tokens from pre-trained backbones in parallel. 
However, despite its memory and runtime benefits, parallel meta-token distillation does not explicitly model joint cross-modal interactions, and how meta-tokens should be leveraged across layers remains unexplored. 
To address this, we propose \textit{MoLT}, a parameter and memory-efficient audio-visual learning framework that enhances cross-modal interaction.
Furthermore, we first explore which transformer layers are truly beneficial for adaptation, revealing that selective layer utilization leads to more effective and efficient learning.

\input{figs/fig3_sub}

%% file: figs/fig2_main.tex
\begin{figure*}[t!]
    \centering
    \includegraphics[width=\textwidth]{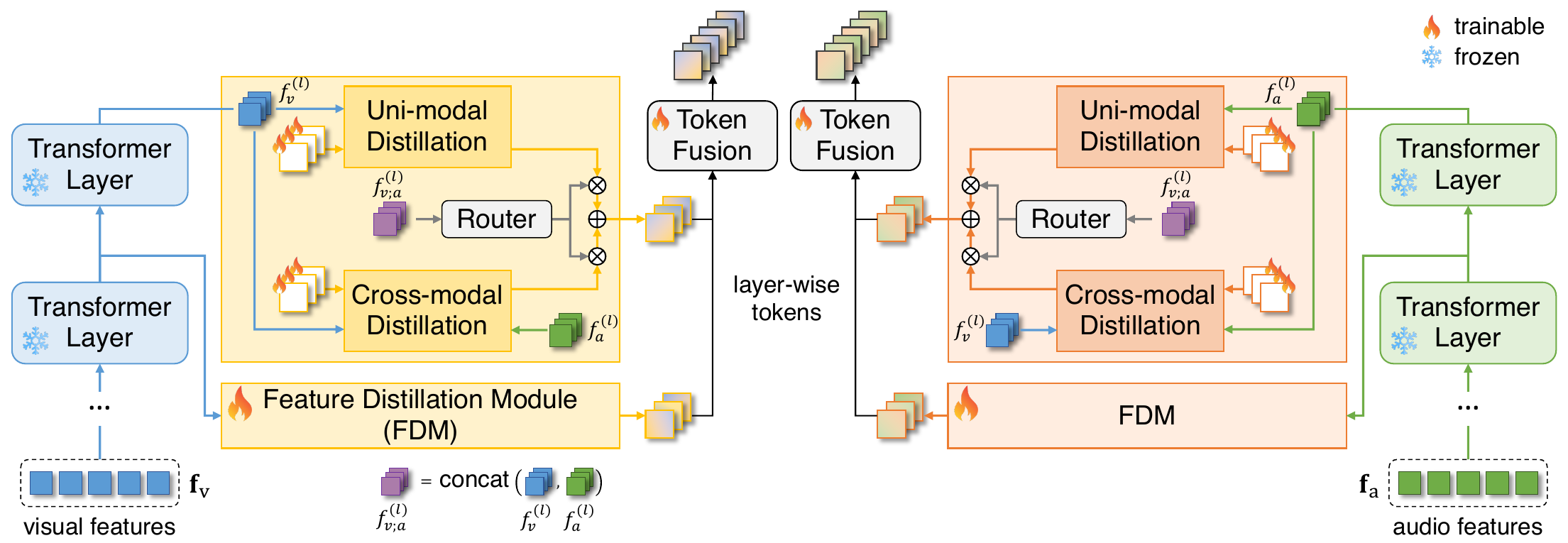}
    \caption{Overview of MoLT framework. The proposed framework consists of three components: (1) pre-trained transformer models, (2) {Feature Distillation Module} that distills cross-modal interactions and modality-specific information from each pre-trained layer into layer-wise latent tokens, and (3) {Token Fusion Module} that assigns importance weights to these tokens and fuses them accordingly.}
    \label{fig:main_fig}
    \vspace{-1em}
\end{figure*}

%% file: figs/fig3_sub.tex
\begin{figure}[t!]
    \centering
    \includegraphics[width=\columnwidth]{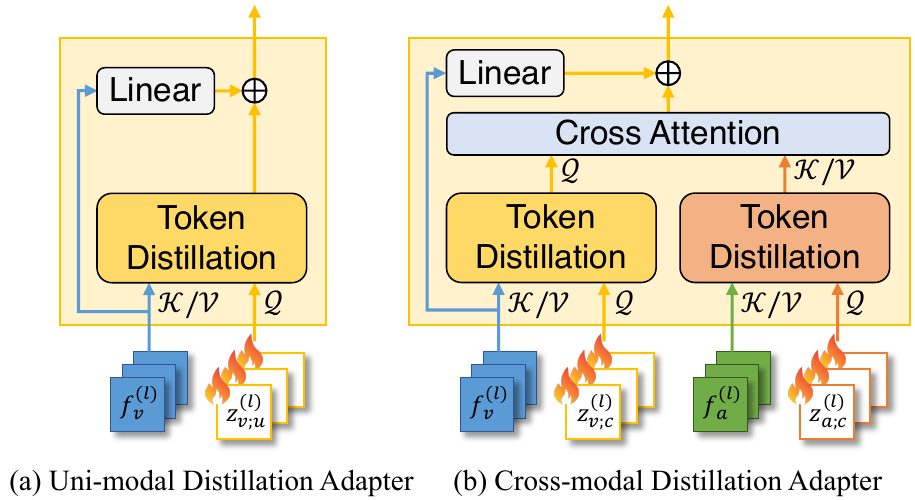}
    \caption{Details of (a) Uni-modal Distillation Adapter, and (b) Cross-modal Distillation Adapter.}
    \label{fig:sub}
    \vspace{-1em}
\end{figure}

%% file: sec/3_method.tex
\section{Method} 
In this section, we elaborate on \textbf{M}ixture \textbf{o}f \textbf{L}ayer-wise \textbf{T}okens (\textbf{MoLT}), an efficient adaptation method for audio-visual learning in terms of both parameters and memory, while maximizing the adaptation performance on diverse audio-visual downstream tasks.
The overall framework of MoLT is illustrated in Figure~\ref{fig:main_fig}.
Our framework consists of pre-trained transformer models (Sec.~\ref{sec:pretrained_transformer}), {Feature Distillation Module} (Sec.~\ref{sec:fdm}) that distills cross-modal interactions and modality-specific information from each pre-trained layer into layer-wise latent tokens, and {Token Fusion Module} that assigns importance weights to these tokens and fuses them accordingly (Sec.~\ref{sec:gating}).
Furthermore, we also detail {Token Orthogonality Regularization} that applied during training (Sec.~\ref{sec:cos_reg}).
This prevents layer-wise tokens from becoming redundant or overly similar, and therefore enables them to learn diverse information required for corresponding downstream tasks. 

\subsection{Pre-trained Transformers}
\label{sec:pretrained_transformer}
Our approach aims to adapt pre-trained transformer models to audio-visual downstream tasks while preserving the original parameters. 
We therefore freeze the audio-visual backbones and train only a small number of additional parameters. 
% For a fair comparison with previous works~\cite{lin2023vision,duan2023cross,cheng2024avmoe,zhou2025mettle}, we adopt Swin-V2-L~\cite{liu2022swin} as the visual backbone and HTS-AT~\cite{chen2022htsat} as the audio backbone.
For a fair comparison with previous works~\cite{lin2023vision,duan2023cross,cheng2024avmoe,zhou2025mettle}, we adopt Swin-V2-L~\cite{liu2022swin} and HTS-AT~\cite{chen2022htsat} as the pre-trained transformer backbones.
% To show the generalizability of our method, we also conduct experiments with other backbone networks, including ViT~\cite{dosovitskiy2021vit}.
Each layer of transformers contains Multi-Head Attention (MHA), a Feed-Forward Network (FFN), and Layer Normalization with residual connections.

\subsection{Feature Distillation Module}
\label{sec:fdm}
As demonstrated in previous works~\cite{cheng2024avmoe}, modality-specific information and cross-modal interaction play complementary roles, necessitating balanced extraction and fusion between the two types of information.
To this end, we propose the Feature Distillation Module (FDM), a Mixture of Experts (MoE) architecture comprising three components: Uni-modal Distillation Adapter (UDA), Cross-modal Distillation Adapter (CDA), and a router.
FDM is applied in a layer-wise manner, but is practically attached only to the later layers of the frozen audio and visual backbones, rather than inserted into every layer as in previous works. 
This depth-aware arrangement is motivated by the characteristics of representations in pre-trained transformers, while significantly reducing parameter and memory overhead. 
We validate this approach with ablations in Sec.~\ref{exp:abl_stage}.

\subsubsection{Uni-modal Distillation Adapter}
For each modality $m \in \{a, v\}$, the uni-modal distillation adapter derives compact latent representations from the $l$-th layer embedding $f_{m}^{(l)} \in \mathbb{R}^{N_{m}^{(l)} \times D_{m}^{(l)}}$ of each backbone.
A set of learnable latent tokens $z^{(l)}_{m;u} \in \mathbb{R}^{N^{(l)}_{m;u} \times D^{(l)}_{m}}$ is used as queries, while the corresponding embeddings $f_{m}^{(l)}$ serve as keys and values within an MHA-based token distillation.
In parallel, the embeddings are projected through a modality-specific linear layer $W_{m;u}^{(l)}$ and added residually to the MHA output, as illustrated in Fig.~\ref{fig:sub}(a).
% This design allows the latent tokens to integrate both attention-refined and linearly projected information from each modality, yielding the distilled representations $\tilde{z}^{(l)}_{m;u}$ at the $l$-th layer for each pre-trained backbone. 
The operation of UDA is formulated as:
\begin{equation}
\tilde{z}^{(l)}_{m;u}
= \text{TD}\!\left(z^{(l)}_{m;u}, f_{m}^{(l)}\right)
+ W^{(l)}_{m;u} f^{(l)}_m,
\end{equation}
where the token distillation $\text{TD}(\cdot)$ is defined as:
\begin{equation}
\scalebox{1}{$\text{TD}\!\left(z^{(l)}_{m;u}, f_{m}^{(l)}\right)
= \sigma\!\left(\frac{z^{(l)}_{m;u} W^Q_m (f_m^{(l)} W^K_m)^{\top}}{\sqrt{d^{(l)}_H}}\right)
f_m^{(l)} W^V_m$}.
\end{equation}
Here, $\sigma$ is softmax, and $W^Q_m$, $W^K_m$, $W^V_m$$\in\mathbb{R}^{D^{(l)}_m\times Hd^{(l)}_H}$ are the projection matrices, where $H$ is the number of heads and $d^{(l)}_H$ is the dimension of each head at $l$-th layer.

\subsubsection{Cross-modal Distillation Adapter}
% It has been demonstrated that models 
Models that rely solely on uni-modal latent representations have been shown to struggle with complex reasoning tasks, such as audio–visual question answering~\cite{zhou2025mettle}. 
To address this limitation, we adopt a cross-modal distillation adapter, as shown in Fig.~\ref{fig:sub}(b), that jointly captures complementary information across modalities and further distills it into a set of additional latent tokens $z^{(l)}_{m;c}$, where $m\in\{a,v\}$.
% The proposed cross-modal distillation adapter yields the output latent tokens $\tilde{z}^{(l)}_{m;c}$ at layer $l$, which can be formulated as follows:
The operation of CDA can be formulated as:
\begin{equation}
\scalebox{0.85}{$\tilde{z}^{(l)}_{m;c}=\text{CA}\!\left(\text{TD}\!\left(z^{(l)}_{m;c},f^{(l)}_m\right),\text{TD}\!\left(z^{(l)}_{\bar{m};c},f^{(l)}_{\bar{m}}\right)\right)+W^{(l)}_{m;c}f^{(l)}_m,$}
\end{equation}
where $\text{CA}(\cdot)$ denotes a cross-attention operation, in which the latent tokens from modality $m$ act as queries and those from the other modality $\bar{m}$ serve as keys and values.
The term $W^{(l)}_{m;c}$ represents a modality-specific linear projection applied to the embedding $f^{(l)}_m$ and added residually to the cross-attention output, enabling stable feature refinement.

%%%%%%%%%%%%%%%%%%%%%%%%%%%%%%%%%%%%
% \input{tabs/main_ave}
\input{tabs/main_avqa}

\input{tabs/main_avs}

\subsubsection{Router}
\label{sec:router}
The router determines which experts to activate based on the characteristics of input features. 
Specifically, concatenated visual and audio embeddings, $f^{(l)}_v$ and $f^{(l)}_a$, are used as inputs $f^{(l)}_{v;a}$ of the router. Following prior work~\cite{fedus2022switch}, the router is implemented as a Multi-Layer Perceptron (MLP). 
The weight for the $i$-th distillation adapter in the FDM is then obtained via a softmax over the router outputs:
\begin{equation}
    w^{(l)}_i = \frac{\text{exp}(R_i(f_{v;a}^{(l)}))}{\sum_{j=1}^{N_\text{UDA}^{(l)}+N_\text{CDA}^{(l)}} \text{exp} (R_j(f_{v;a}^{(l)}))},
\end{equation}
where $R_i(f_{v;a})$ denotes the output of the router for the $i$-th distillation adapter, and $N_{\text{UDA}}^{(l)}, N_{\text{CDA}}^{(l)}$ denote the numbers of uni- and cross-modal distillation adapters, respectively.
These weights are applied to the adapter outputs to compute a weighted sum, yielding the final layer-wise latent tokens $\tilde{z}_m^{(l)}$ for the corresponding transformer layer as follows:
\begin{equation}
\tilde{z}^{(l)}_m=\sum_{i=1}^{N^{(l)}_{\text{UDA}}}w^{(l)}_{u_i}\tilde{z}^{(l)}_{m;u_i}+\sum_{j=1}^{N^{(l)}_{\text{CDA}}}w^{(l)}_{c_j}\tilde{z}^{(l)}_{m;c_j}.
\end{equation}

\subsection{Token Fusion Module}
\label{sec:gating}
Since each transformer layer captures different levels of semantics, it is important to fuse the layer-wise information relevant to the downstream task. To this end, we employ a modality-specific Token Fusion Module (TFM).
It takes the concatenated layer-wise latent tokens $\tilde{\textbf{Z}}_m = [\tilde{z}_m^{(l')},\tilde{z}_m^{(l'+1)},\cdots,\tilde{z}_m^{(L)}]$ as input and produces the fused representation $\hat{\textbf{Z}}_m$ according to their relative importance. The TFM is implemented as an MLP followed by a softmax to compute layer-wise fusion weights:
\begin{equation}
    \alpha^{(l)}_m=\frac{\text{exp}(\text{MLP}(\tilde{z}_{m}^{(l)}))}{\sum_{k=l'}^{L} \text{exp} (\text{MLP}(\tilde{z}_m^{(k)}))},
\end{equation}
and the final fused representation is obtained as
\begin{equation}
    \hat{\textbf{Z}}_m=\sum_{l=l'}^{L}\alpha_m^{(l)}\tilde{z}_m^{(l)},
\end{equation}
where $l'$ and $L$ denote the starting and final layers of the backbone, respectively. The resulting fused representation is used as the modality embedding for the final prediction in downstream tasks.

\subsection{Token Orthogonality Regularization}
\label{sec:cos_reg}
When extracting layer-wise latent tokens through FDMs, there is a risk that the resulting tokens become overly redundant or even identical.
Such redundancy limits the diversity of semantic information captured by the latent tokens, potentially weakening the effectiveness of adaptation.
To address this issue, we apply Token Orthogonality Regularization (TOR) during training. Given the L2-normalized concatenated sequence of layer-wise latent tokens $\tilde{\textbf{Z}}_m=\{\tilde{z}_{m,i}\}^{N}_{i=1}$, where $N$ is the total number of layer-wise tokens, the TOR loss is computed as follows:
\begin{equation}
    \mathcal{L}_{\text{TOR}}=||\textbf{C}-\textbf{I}_N||^2,\ \ \ \text{where}\ \ C_{ij}=\frac{\tilde{z}_{m,i}\cdot\tilde{z}_{m,j}}{\|\tilde{z}_{m,i}\| \|\tilde{z}_{m,j}\|},
\end{equation}
where $\textbf{I}_N$ denotes $N \times N$ identity matrix. 
This constraint encourages the FDMs to extract mutually distinctive yet complementary representations, which in turn enhances the model’s adaptation capability.

%% file: tabs/main_avqa.tex
\begin{table*}[t!]
\centering
\setlength{\tabcolsep}{6pt}
\caption{\textbf{Audio-Visual Question Answering}. Adaptation performance comparison on the test set of the MUSIC-AVQA dataset.
The overall average accuracy with accuracy on each type of question, audio question (AQ), visual question (VQ), and audio-visual question (AVQ), is reported.
$^{\dagger}$marks results reported in previous work~\cite{cheng2024avmoe}. $^*$reproduced in our environment.}
\small
% \begin{tabularx}{\textwidth}{l>{\centering\arraybackslash}X>{\centering\arraybackslash}X>{\centering\arraybackslash}X>{\centering\arraybackslash}X>{\centering\arraybackslash}X>{\centering\arraybackslash}X>{\centering\arraybackslash}X>{\centering\arraybackslash}X}
% \begin{adjustbox}{width=0.85\textwidth}
\begin{adjustbox}{width=\textwidth}
\begin{tabular}{lccccc|cccc}
\toprule[1pt]
%        & Visual & Audio & Trainable & Total & \multicolumn{4}{c}{Modality Metric}  \\
% Method & Encoder & Encoder & Params($\%$)$\downarrow$ & Params(M)$\downarrow$ & AQ & VQ & AVQ & Avg.\\
\multirow{2}{*}{Method} & Visual & Audio & Trainable & Total & Memory & \multirow{2}{*}{AQ$\uparrow$} & \multirow{2}{*}{VQ$\uparrow$} & \multirow{2}{*}{AVQ$\uparrow$} & \multirow{2}{*}{Avg.$\uparrow$} \\
  & Encoder & Encoder & Params($\%$)$\downarrow$ & Params(M)$\downarrow$ & (GB)$\downarrow$& & &  &\\
\midrule
% \textbf{\textit{with pre-train}} \\
AVSD~\cite{schwartz2019simple}          &   VGG-19      &   VGG-like    &  -  &  -  &  - & 68.5  &  70.8  &  65.5  &  67.4  \\
Pano-AVQA~\cite{yun2021pano}     &   Faster RCNN &   VGG-like    &  -  &  -  &  - &   70.7  &  72.6  &  66.6  &  68.9  \\
ST-AVQA~\cite{li2022learning}       &   ResNet-18   &   VGG-like    &  11.2  &  94.4  &  -  &  74.1  &  74.0  &  69.5  &  71.5  \\
LAVisH~\cite{lin2023vision}        &   \multicolumn{2}{c}{Swin-V2-L (shared)}&  8.4  &  249.8  &  10.8 &   75.7  &  80.4  &  70.4  &  74.0  \\
\textbf{MoLT(ours)}        
              &    \multicolumn{2}{c}{Swin-V2-L (shared)} & 19.5   &  366.2    &  1.7 &   \textbf{75.8}  &  \textbf{81.3}  &  \textbf{71.5}   &  \textbf{74.9}    \\
\midrule
LAVisH$^{\dagger}$~\cite{lin2023vision}       &    Swin-V2-L      &     HTS-AT    &  28.7  &  367.4  &  - &   75.4  &  79.6   &  70.1  & 73.6 \\
DG-SCT~\cite{duan2023cross}        &    Swin-V2-L      &     HTS-AT    &  50.0  &  520.2  &  11.3 &   77.4  &  81.9   &  70.7  & 74.8 \\
AVMoE~\cite{cheng2024avmoe}         &    Swin-V2-L      &     HTS-AT    &  42.7  &  456.6  &  11.4 &   \textbf{77.6}  &  82.7   &  71.9  & 75.7 \\
Mettle$^*$~\cite{zhou2025mettle}        &     Swin-V2-L      &     HTS-AT    & 29.4     &  452.2 &  3.1 &   74.9  & 79.0 & 69.4 & 72.9 \\  % 3.1 GB
\textbf{MoLT(ours)}        
              &    Swin-V2-L      &     HTS-AT    & 25.9  &  441.4    &  2.0 &  77.4  &  \textbf{83.1}  &  \textbf{72.3}   &  \textbf{76.1}    \\

\bottomrule[1pt]
% \end{tabularx}
\end{tabular}
\end{adjustbox}
\label{tab:main_avqa}
\end{table*}

%% file: tabs/main_avs.tex
\begin{table*}[t!]
\centering
\setlength{\tabcolsep}{8pt}
\caption{\textbf{Audio-Visual Segmentation}. Adaptation performance comparison under the single-source (S4) and multi-source (MS3) settings of the AVSBench dataset.}
\small
% \begin{tabularx}{\textwidth}{l>{\centering\arraybackslash}X>{\centering\arraybackslash}X>{\centering\arraybackslash}X>{\centering\arraybackslash}X>{\centering\arraybackslash}X>{\centering\arraybackslash}X>{\centering\arraybackslash}X>{\centering\arraybackslash}X>{\centering\arraybackslash}X}
% \begin{adjustbox}{width=\textwidth}
\begin{adjustbox}{width=\textwidth}
\begin{tabular}{lccccc|cc|cc}
\toprule[1pt]
\multirow{2}{*}{Method} & Visual & Audio & Trainable & Total & Memory   & \multicolumn{2}{c}{S4}  &  \multicolumn{2}{c}{MS3}  \\
  & Encoder & Encoder & Params($\%$)$\downarrow$ & Params(M)$\downarrow$ & (GB)$\downarrow$ & $\mathcal{M}_{\mathcal{J}}$$\uparrow$ & $\mathcal{M}_{\mathcal{F}}$$\uparrow$ & $\mathcal{M}_{\mathcal{J}}$$\uparrow$ & $\mathcal{M}_{\mathcal{F}}$$\uparrow$ \\
\midrule
% \textbf{\textit{with pre-train}} \\
AVS~\cite{zhou2022avs}           &    PVT-V2         &     VGGish    &    58.7  &  174.5  &  4.8   &  78.7  &  87.9    &   54.0  &  64.5 \\
% \midrule
LAVisH~\cite{lin2023vision}        &    \multicolumn{2}{c}{Swin-V2-L (shared)} &    14.0    &  266.4  &  5.9  &  80.1  &  88.0  & 49.8  &  60.3 \\
Mettle~\cite{zhou2025mettle}        &    \multicolumn{2}{c}{Swin-V2-L (shared)} &    11.1    &  257.2  &  2.7  &  79.9  &   -    & 53.7  &   -    \\
\textbf{MoLT(ours)}        
              &    \multicolumn{2}{c}{Swin-V2-L (shared)} &     8.3     &  313.1  &  3.6   &  \textbf{80.2}  &  \textbf{89.5}   &  \textbf{55.1}  &  \textbf{68.2} \\
\midrule
LAVisH~\cite{lin2023vision}        &    Swin-V2-L      &     HTS-AT    &     47.6  &  389.7  &  -    &  78.0  & 87.0   & 49.1  &  59.9 \\
DG-SCT~\cite{duan2023cross}        &    Swin-V2-L      &     HTS-AT    &     61.5  &  594.8  &  9.0  &  80.9  & 89.2   & 53.5  &  64.2 \\
AVMoE~\cite{cheng2024avmoe}         &    Swin-V2-L      &     HTS-AT    &     54.4  &  501.2  &  9.5  &  \textbf{81.1}  & 89.7   & 54.5  &  68.7 \\
Mettle~\cite{zhou2025mettle}        &    Swin-V2-L      &     HTS-AT    &     21.6  &  291.7  &  2.9  &  80.7  &    -   & 55.1  &   -   \\
\textbf{MoLT(ours)}        
              &    Swin-V2-L      &     HTS-AT    & 6.3 &  339.5    &  3.7  &  80.8 &  \textbf{89.9}  &  \textbf{57.1}   & \textbf{70.3}  \\

\bottomrule[1pt]
% \end{tabularx}
\end{tabular}
% \end{adjustbox}
\label{tab:main_avs}
\end{adjustbox}
\end{table*}

%% file: sec/4_exp.tex
\input{tabs/main_ave}

\section{Experiments}

\subsection{Experimental Setup}

\noindent\textbf{Datasets and Metrics.}
We evaluate our method on three audio–visual learning tasks: Audio–Visual Question Answering (AVQA), Audio–Visual Segmentation (AVS), and Audio–Visual Event Localization (AVE). 
For AVQA, we use the MUSIC-AVQA dataset~\cite{li2022learning}, which provides 45,867 question–answer pairs for 9,288 musical instrument video clips, and we measure accuracy following prior works.
The AVS task is conducted on the AVSBench dataset~\cite{zhou2022avs}, which includes 4,932 single-source (S4) and 404 multi-source (MS3) videos covering 23 classes; we employ the Jaccard index $\mathcal{M}_{\mathcal{J}}$ [12] and F-score $\mathcal{M}_{\mathcal{F}}$ as evaluation metrics, which respectively quantify the region-level overlap and boundary precision of the predicted masks. 
The AVE dataset~\cite{tian2018audio} contains 4,143 videos across 28 event categories, and we report segment-level accuracy for evaluation. 

\noindent\textbf{Implementation Details.}
Our framework is built upon large-scale pre-trained transformers, where Swin-V2-L~\cite{liu2022swin} serves as the visual backbone and HTS-AT~\cite{chen2022htsat} as the audio counterpart. To explore the transferability from vision domain to audio-visual learning, we further experiment with shared Swin-V2-L backbones, originally trained on visual data only. 
% we further experiment with shared backbones such as ViT-L-16~\cite{dosovitskiy2021vit} and Swin-V2-L, 
Task-specific decoders are subsequently integrated to generate predictions, following the setups of~\cite{zhou2025mettle,cheng2024avmoe,duan2023cross}. 
% Detailed configurations, including batch size, learning rate, training epochs, and task-dependent hyperparameters, are described in the supplementary material. 
All experiments are performed on a single NVIDIA RTX 4090 GPU.

%%%%%%%%%%%%%%%%%%%%%%%%%%%%%%%%%%%%%%%%%%%%%%%%%%%%%%%%%%%%%
\subsection{Main Results}

\subsubsection{{Audio-Visual Question Answering}}
AVQA requires complex reasoning and serves as a benchmark to evaluate both the generalizability and reasoning capability of parameter- and memory-efficient models.
The task includes three question categories, audio (AQ), visual (VQ), and audio-visual (AVQ) that demand comprehensive multimodal understanding and spatio-temporal reasoning.
As shown in Tab.~\ref{tab:main_avqa}, MoLT achieves strong performance across all categories with a small number of trainable parameters, outperforming previous methods under both shared and separate encoder settings.
These results demonstrate that MoLT effectively performs cross-modal reasoning by distilling both uni-modal and cross-modal representations into layer-wise latent tokens.
By utilizing only the most informative layers, MoLT maintains memory efficiency while preserving strong reasoning capability across complex multimodal question types.

\subsubsection{{Audio-Visual Segmentation}}
To demonstrate the effectiveness of the proposed method on the pixel-wise dense prediction task, we compare MoLT with previous adaptation approaches on AVS task.
As shown in Tab.~\ref{tab:main_avs}, our proposed MoLT achieves superior performance to previous methods on the AVSBench dataset under both S4 and MS3 settings, regardless of whether employing separate or weight-shared transformer backbones.
In particular, on MS3, both $\mathcal{M}_{\mathcal{J}}$ and $\mathcal{M}_{\mathcal{F}}$ show significant improvements over previous works, demonstrating the robustness of our method in complex multi-sound source scenarios.
Moreover, despite having the smallest proportion of trainable parameters, MoLT achieves these superior results, highlighting its efficiency and effectiveness.

\subsubsection{Audio-Visual Event Localization}
AVE requires fine-grained cross-modal correspondence and temporal alignment. 
As summarized in Tab.~\ref{tab:main_ave}, MoLT achieves the highest accuracy while using only 6.2$\%$ trainable parameters and 0.7 GB GPU memory. 
Although accuracy of the Swin-V2-L shared-backbone setting trails the previous works, MoLT still shows strengths in terms of parameters and memory.
Together, the results confirm that focusing adaptation on late layers and mixing uni-modal and cross-modal token distillation yields state-of-the-art AVE performance under constrained compute, with the best accuracy–efficiency trade-off across configurations.

%%%%%%%%%%%%%%%%%%%%%%%%%%%%%%%%%%%%%%%%%%%%%%%%%%%%%%%%%%%%%%%%
\input{figs/fig_cos_layer}
\input{tabs/abl_position}

\subsection{Ablation Studies}

\subsubsection{Optimal Adaptation Position}
\label{exp:abl_stage}

To determine the most effective adaptation position for pre-trained transformers in audio-visual learning, we conduct experiments on various adaptation positions.
We first analyzed the frozen transformer backbones by comparing the cosine similarities between the input and output of each layer to define these positions.
Based on this analysis (as shown in Fig.~\ref{fig:sup_cos_layer}), we define the first four layers - which exhibit the lowest cosine similarity, indicating the largest change in representation - as the \textit{early} stage.
Then the final two layers, which show relatively smaller changes and encode high-level semantics, are defined as the \textit{late} stage, and the remaining layers in between are regarded as the \textit{middle} stage.
Unlike previous works that insert adapters into every transformer layer, we selectively adapt only late layers in the proposed MoLT framework.
As shown in Tab.~\ref{exp:abl_stage}, this design achieves the strongest performance, while remaining efficient in terms of parameters and memory. 
We attribute the effectiveness of late-layer adaptation to two factors. 
First, late transformer layers predominantly encode high-level semantics that are directly relevant for downstream prediction, so adjusting the representations of these layers is more effective. 
Second, according to our analysis, representations undergo the largest transitions across the early transformer layers.
When adaptation is applied inaccurately at this stage, the resulting errors can accumulate and propagate through subsequent layers, finally degrading the adaptation performance.

\input{tabs/abl_num_token}
\input{tabs/abl_num_expert}
\input{figs/fig4_reg}

\subsubsection{Components of FDM}

\noindent\textbf{Number of latent tokens for each modality distillation.}
We conduct ablations on the number of latent tokens for each modality in the distillation adapters of FDM.
As shown in Tab.~\ref{tab:abl_num_token}, we observe monotonic performance gains on AVE and AVQA as the number of latent tokens per modality increases, whereas the performance on AVS peaks at a moderate size. 
In the proposed MoLT framework, layer-wise tokens summarize high-level, global semantics from late layers so that increasing their diversity raises the representational capacity and benefits tasks that depend on holistic evidence. 
For AVS, however, accurate masks rely on spatially precise, pixel-level cues.
Adding too many late-layer tokens may introduce conflicting global signals, and the softmax-based token fusion module tends to spread weights across more tokens, leading to over-smoothing and weaker localization. 
Consequently, a moderate number of tokens provides the best trade-off for dense prediction, while larger token sets help classification and reasoning tasks.

\noindent\textbf{Number of distillation adapters in FDM.} 
FDM can adopt multiple uni-modal and cross-modal distillation adapters.
Tab.~\ref{tab:abl_num_expert} shows the performance depending on the number of each distillation adapter.
Increasing the number of either adapter improves the average performance, with the best overall results obtained when using two uni-modal and two cross-modal adapters.
Adding more cross-modal adapters yields the largest gain on AQ, suggesting that visual context helps disambiguate acoustic events. Increasing uni-modal adapters primarily benefits VQ, indicating the value of modality-specific detail. 
AVQ peaks only when both capacities are present, highlighting the complementarity between modality-specific and cross-modal experts.
From this results, we can infer that each distillation adapter effectively distills complementary information useful for downstream tasks from each transformer layer.

\input{tabs/abl_fusion}

\subsubsection{Fusion of Layer-wise Latent Tokens}
We conduct an ablation study to determine an effective method for fusing the layer-wise latent tokens extracted via FDM before passing through the prediction head.
We compare three approaches: (1) a simple average pooling strategy~\cite{zhou2025mettle}, (2) a weighted sum where each token's weight is a distinct learnable parameter acting as a gate, and (3) an MLP-based token fusion.
As shown in Tab.~\ref{tab:abl_fusion}, the MLP-based token fusion yields the best performance.
This indicates that capturing the relative importance among tokens in an input-adaptive manner is crucial for effective fusion.

\subsubsection{Token Orthogonality Regularization}

We also analyze the effect of token orthogonality regularization.
As shown in Tab.~\ref{tab:abl_reg}, applying TOR consistently improves performance across downstream tasks.
To understand its mechanism, we visualize the cosine similarity map between the latent tokens and their weights generated by TFM.
Fig.~\ref{fig:reg} (a) shows that without TOR the latent tokens in both the audio and visual branches tend to collapse, becoming almost identical across layers.
Correspondingly, Fig.~\ref{fig:reg} (b) indicates that the TFM assigns nearly uniform weights to these tokens.
In contrast, applying TOR significantly reduces redundancy between tokens, allowing the TFM to assign more differentiated, input-sensitive weights.
Taken together, TOR mitigates informational overlap among latent tokens and thereby enables the TFM to better capture their relative importance, yielding more effective fusion.

\input{tabs/abl_reg}

%% file: tabs/main_ave.tex
\begin{table*}[t!]
\centering
\setlength{\tabcolsep}{8pt}
\caption{\textbf{Audio-Visual Event Localization}. Performance comparison on the test set of AVE dataset. $^*$reproduced in our environment.}
\small
% \begin{tabularx}{\textwidth}{l>{\centering\arraybackslash}X>{\centering\arraybackslash}X>{\centering\arraybackslash}X>{\centering\arraybackslash}X>{\centering\arraybackslash}X>{\centering\arraybackslash}X>{\centering\arraybackslash}X>{\centering\arraybackslash}X>{\centering\arraybackslash}X}
% \begin{adjustbox}{width=\textwidth}
\begin{adjustbox}{width=\textwidth}
\begin{tabular}{lccccccc|c}
\toprule[1pt]
% & \multicolumn{2}{c}{Encoder} & \multicolumn{2}{c}{PT Dataset} & \multicolumn{2}{c}{Params} & Memory & Runtime & Acc  \\
% \cmidrule(lr){2-3} \cmidrule(lr){4-5} \cmidrule(lr){6-7}
% Method & Visual & Audio &  Visual  & Audio & Trainable($\%$) & Total(M) & (GB) & (ms)  & $\uparrow$  \\
\multirow{2}{*}{Method} & Visual & Audio &  Visual  & Audio & Trainable & Total & Memory & \multirow{2}{*}{Acc.$\uparrow$}  \\
  & Encoder & Encoder &  PT Dataset  & PT Dataset & Params($\%$)$\downarrow$ & Params(M)$\downarrow$ & (GB)$\downarrow$ &   \\
\midrule
% \textbf{\textit{with pre-train}} \\
AVEL~\cite{tian2018audio}          &    ResNet-152    &   VGGish  &   ImageNet    &  AudioSet   &  2.7  &  136.0  &  -  &  74.0 \\
AVSDN~\cite{lin2019dual}         &    ResNet-152    &   VGGish  &   ImageNet    &  AudioSet   &  5.7  &  140.3  &  -  &  75.4 \\
CMRAN~\cite{xu2020cross}         &    ResNet-152    &   VGGish  &   ImageNet    &  AudioSet   &  10.7 &  148.2  &  -  &  78.3 \\
MM-Pyramid~\cite{yu2022mm}    &    ResNet-152    &   VGGish  &   ImageNet    &  AudioSet   &  25.0 &  176.3  &  -  &  77.8 \\
CMBS~\cite{xia2022crossmodal}          &    ResNet-152    &   VGGish  &   ImageNet    &  AudioSet   &  6.6  &  216.7  &  -  &  79.7 \\
\midrule
% LAVisH        &    \multicolumn{2}{c}{ViT-L-16 (shared)}  &   ImageNet    &  \xmark   &  4.3   &  340.1  &  19.2  &  78.1 \\
% AVMoE         &    \multicolumn{2}{c}{ViT-L-16 (shared)}  &   ImageNet    &  \xmark   &  32.6  &  483.1  &  19.5  &  79.2 \\
% MA-AVT        &    \multicolumn{2}{c}{ViT-L-16 (shared)}  &   ImageNet    &  \xmark   &  3.7   &  338.4  &  7.2   &  79.6 \\
% Mettle        &    \multicolumn{2}{c}{ViT-L-16 (shared)}  &   ImageNet    &  \xmark   &  38.8  &  532.2  &  1.7   &  80.6 \\
% \textbf{MoLT(ours)}        
%               &    \multicolumn{2}{c}{ViT-L-16 (shared)}  &   ImageNet    &  \xmark   &  -     &  -  &  -  &  \lhk{76.3} \\
% \midrule
LAVisH~\cite{lin2023vision}        &    \multicolumn{2}{c}{Swin-V2-L (shared)} &   ImageNet    &  \xmark   &  2.7    &  238.8  &  15.4  &  81.1 \\
AVMoE~\cite{cheng2024avmoe}         &    \multicolumn{2}{c}{Swin-V2-L (shared)} &   ImageNet    &  \xmark   &  39.4   &  374.4  &  16.8  &  81.5 \\
Mettle~\cite{zhou2025mettle}        &    \multicolumn{2}{c}{Swin-V2-L (shared)} &   ImageNet    &  \xmark   &  37.2   &  364.4  &  1.8   &  82.3 \\
Mettle$^*$~\cite{zhou2025mettle}        &    \multicolumn{2}{c}{Swin-V2-L (shared)} &   ImageNet    &  \xmark   &  37.0 &  363.3  &  1.3   &  78.0 \\
\textbf{MoLT(ours)}        
              &    \multicolumn{2}{c}{Swin-V2-L (shared)} &   ImageNet    &  \xmark   &  3.0 &  248.5  &  0.6  &  79.7 \\
\midrule
LAVisH~\cite{lin2023vision}        &    Swin-V2-L      &     HTS-AT    &   ImageNet    &  AudioSet   &  30.6  &  374.9  &  11.9  &  78.6 \\
DG-SCT~\cite{duan2023cross}        &    Swin-V2-L      &     HTS-AT    &   ImageNet    &  AudioSet   &  43.6  &  461.3  &  13.1  &  82.2 \\
AVMoE~\cite{cheng2024avmoe}         &    Swin-V2-L      &     HTS-AT    &   ImageNet    &  AudioSet   &  34.9  &  404.0  &  12.1  &  82.6 \\
Mettle~\cite{zhou2025mettle}        &    Swin-V2-L      &     HTS-AT    &   ImageNet    &  AudioSet   &  23.0  &  338.0  &  1.5   &  83.3 \\
Mettle$^*$~\cite{zhou2025mettle}        &    Swin-V2-L      &     HTS-AT    &   ImageNet    &  AudioSet   &  30.6  &  374.9  &  1.3  &  81.5 \\
\textbf{MoLT(ours)}        
              &    Swin-V2-L      &     HTS-AT    &   ImageNet    &  AudioSet   &  6.2     &  290.3  &  0.7  &  \textbf{83.5}  \\
\bottomrule[1pt]
% \end{tabularx}
\end{tabular}
% \end{adjustbox}
\label{tab:main_ave}
\end{adjustbox}
\end{table*}

%% file: figs/fig_cos_layer.tex
% \twocolumn[{%
% \begin{center}
% \includegraphics[width=0.9\textwidth]{figs/sup/sup_cos_sim_layer.pdf}
% \captionof{figure}{Cosine similarity between input and output of each transformer layer for audio and visual branches on AVE task.} % \usepackage{caption} 필요
% \label{fig:sup_cos_layer}
% \end{center}
% \vspace{-0.8em}
% }]

\begin{figure*}[t]
    \centering
    \includegraphics[width=0.75\textwidth]{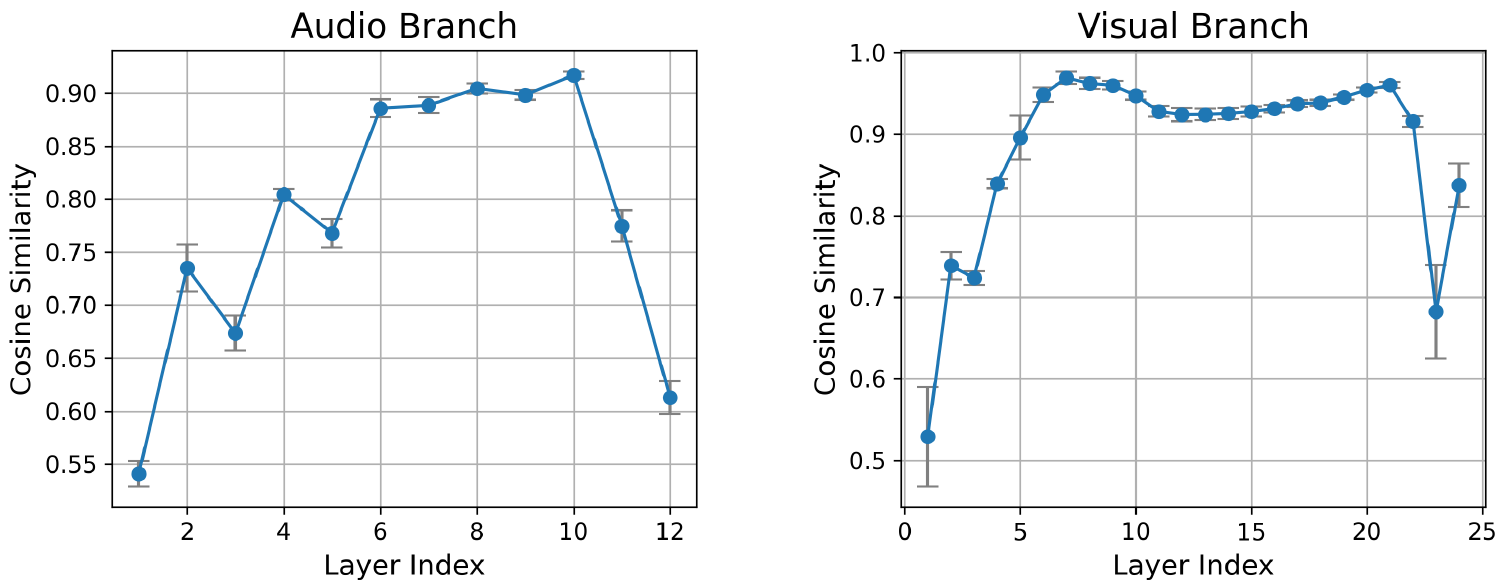}
    \caption{Cosine similarity between input and output of each transformer layer for audio and visual branches on the AVE task.}
    \label{fig:sup_cos_layer}
    \vspace{-1em}
\end{figure*}

%% file: tabs/abl_position.tex
\begin{table}[t!]
\centering
% \vspace{-0.5em}
% \renewcommand{\arraystretch}{1.1}
\caption{Performance comparison across different adaptation positions in pre-trained transformers.}
\setlength{\tabcolsep}{3pt}
\small
% \begin{tabular}{ccc|c|cccc}
\begin{tabularx}{.98\columnwidth}{ccc|>{\centering\arraybackslash}X>{\centering\arraybackslash}X>{\centering\arraybackslash}X|>{\centering\arraybackslash}X>{\centering\arraybackslash}X>{\centering\arraybackslash}X}
\toprule
% \multicolumn{2}{c|}{\textbf{Num. FDMs}}&
\multicolumn{3}{c|}{Adaptation Stage} 
& \multicolumn{3}{c|}{AVQA}
& \multicolumn{3}{c}{AVE} \\
% \cmidrule(lr){1-3} \cmidrule(lr){4} \cmidrule(lr){5-8}
% \cmidrule(lr){3} \cmidrule(lr){4-7}
Early & Mid & Late & Param. & Mem. & Acc. & Param. & Mem. & Acc. \\
\midrule
\cmark  & \cmark    & \cmark     & 35.1 & 3.9 & 75.2     & 14.7 & 1.0 & 81.8\\
\midrule
\cmark  & \cmark    & \xmark     & 21.2 & 3.4 & 73.4     & 9.7 & 0.9 & 80.6 \\
\cmark  & \xmark    & \cmark    & 23.4 & 3.0 & 74.6      & 8.0 & 0.8 &  81.2 \\
\xmark  & \cmark    & \cmark    & 33.6 & 2.6 & 75.6     & 13.2 & 0.8 &  82.1  \\
\midrule
\cmark  & \xmark    & \xmark    & 3.4 & 2.5 & 68.7      & 2.1 & 0.7 &  63.1  \\
\xmark  & \cmark    & \xmark    & 19.0 & 2.2 & 73.8     & 8.0 & 0.7 &  80.1  \\
\xmark  & \xmark    & \cmark    & 25.9 & 2.0 & \textbf{76.1}       & 6.3 & 0.7 & \textbf{83.5}  \\
\bottomrule
\end{tabularx}
\label{tab:abl_position}
\vspace{-1.2em}
\end{table}

%% file: tabs/abl_num_token.tex
\begin{table}[t!]
\centering
% \vspace{-0.5em}
% \renewcommand{\arraystretch}{1.1}
\caption{Ablation studies on the effects of the number of latent tokens for each modality.}
\small
% \begin{tabular}{cc|c|cc|cc|c}
\begin{tabularx}{\columnwidth}{cc|>{\centering\arraybackslash}X|cc|cc|>{\centering\arraybackslash}X}
\toprule
\multicolumn{2}{c|}{$\#$ Tokens}&
\multicolumn{1}{c|}{AVQA} &
\multicolumn{2}{c|}{AVS-S4} &
\multicolumn{2}{c|}{AVS-MS3} &
\multicolumn{1}{c}{AVE}  \\

A & V &
Acc. &
{\scriptsize $\mathcal{M}_{\mathcal{J}}$} & {\scriptsize $\mathcal{M}_{\mathcal{F}}$} &
{\scriptsize $\mathcal{M}_{\mathcal{J}}$} & {\scriptsize $\mathcal{M}_{\mathcal{F}}$} &
Acc. \\

\midrule
2 & 2       & 74.9       & 80.7 & 89.8 & 55.5 & 68.6        & 81.8 \\
4 & 4      & 75.3         & \textbf{80.8} & \textbf{89.9} & \textbf{57.1} & \textbf{70.3}         & 82.1 \\
6 & 6      & 75.5       & 80.4 & 89.5 & 55.5 & 69.9              & 82.5 \\
8 & 8     & \textbf{76.1}      & 80.6 & 89.7 & 55.7 & 69.9          & \textbf{83.5} \\
\bottomrule
\end{tabularx}
\label{tab:abl_num_token}
\vspace{-0.2em}
\end{table}

% \begin{table*}[t!]
% \centering
% % \renewcommand{\arraystretch}{1.1}
% \caption{Effects of the number of latent tokens for each modality.}
% \setlength{\tabcolsep}{4.5pt}
% \begin{tabular}{cc|c|cc|cc|cccc}
% \toprule
% \multicolumn{2}{c|}{Num. Tokens}&
% \multicolumn{1}{c|}{AVE} &
% \multicolumn{2}{c|}{AVS (S4)} &
% \multicolumn{2}{c|}{AVS (MS3)} &
% \multicolumn{4}{c}{AVQA}  \\

% A & V &
% Acc. &
% $\mathcal{M}_{\mathcal{J}}$ & $\mathcal{M}_{\mathcal{F}}$ &
% $\mathcal{M}_{\mathcal{J}}$ & $\mathcal{M}_{\mathcal{F}}$ &
% AQ & VQ & AVQ & Avg. \\

% \midrule
% 2 & 2 & - & - & - & - & - & 76.4 & 82.4 & 70.8 & 74.9 \\
% 4 & 4 & - & - & - & - & - & 77.8 & 81.8 & 71.8 & 75.5 \\
% 6 & 6 & - & - & - & - & - & 77.4 & 81.8 & 71.5 & 75.3 \\
% 8 & 8 & - & - & - & - & - & 77.4  &  \textbf{83.1}  &  \textbf{72.3}   &  \textbf{76.1} \\
% \bottomrule
% \end{tabular}

% \label{tab:abl_num_expert}
% \end{table*}

%% file: tabs/abl_num_expert.tex
\begin{table}[t!]
\centering
\caption{Ablation results for utilizing different numbers of cross-modal and uni-modal distillation adapters in FDM.}
\small
% \begin{tabular}{cc|cc|cc|cccc}
\setlength{\tabcolsep}{3pt} % 기본은 6pt 
\begin{tabularx}{\columnwidth}{
>{\centering\arraybackslash}X>{\centering\arraybackslash}X|
% >{\centering\arraybackslash}X>{\centering\arraybackslash}X|
% >{\centering\arraybackslash}X>{\centering\arraybackslash}X|
>{\centering\arraybackslash}X>{\centering\arraybackslash}X
>{\centering\arraybackslash}X>{\centering\arraybackslash}X}
\toprule
% \multicolumn{2}{c|}{$\#$ FDM}&
% \multicolumn{2}{c|}{}&
% \multicolumn{2}{c|}{AVS-S4} &
% \multicolumn{2}{c|}{AVS-MS3} &
% \multicolumn{4}{c}{AVQA Acc.}  \\

$\#$UFA & $\#$CFA &
% {\scriptsize $\mathcal{M}_{\mathcal{J}}$} & {\scriptsize $\mathcal{M}_{\mathcal{F}}$} &
% {\scriptsize $\mathcal{M}_{\mathcal{J}}$} & {\scriptsize $\mathcal{M}_{\mathcal{F}}$} &
% ${\mathcal{J}}$ & ${\mathcal{F}}$ &
% ${\mathcal{J}}$ & ${\mathcal{F}}$ &
AQ & VQ & AVQ & Avg. \\

\midrule
1 & 1 &  76.1 & 81.2 & 70.9 & 74.5 \\
2 & 1 &  77.7 & 81.8 & 71.2 & 75.2 \\
1 & 2 &  \textbf{78.3} & 81.2 & 71.5 & 75.3 \\
2 & 2 &  77.4 & \textbf{83.1} & \textbf{72.3} & \textbf{76.1} \\
\bottomrule
\end{tabularx}
\label{tab:abl_num_expert}
\vspace{-0.2em}
\end{table}

% \begin{table}[t!]
% \centering
% % \renewcommand{\arraystretch}{1.1}
% \caption{Performance comparison across different numbers of cross-modal and uni-modal distillation adapters in FDM. "Cross" and "Uni" denote cross-modal and uni-modal respectively. }
% % \setlength{\tabcolsep}{4.5pt}
% \small
% % \begin{tabular}{cc|cc|cc|cccc}
% \setlength{\tabcolsep}{3pt} % 기본은 6pt 
% \begin{tabularx}{\columnwidth}{>{\centering\arraybackslash}X>{\centering\arraybackslash}X|>{\centering\arraybackslash}X>{\centering\arraybackslash}X|>{\centering\arraybackslash}X>{\centering\arraybackslash}X|>{\centering\arraybackslash}X>{\centering\arraybackslash}X>{\centering\arraybackslash}X>{\centering\arraybackslash}X}
% \toprule
% % \multicolumn{2}{c|}{$\#$ FDM}&
% \multicolumn{2}{c|}{}&
% \multicolumn{2}{c|}{AVS-S4} &
% \multicolumn{2}{c|}{AVS-MS3} &
% \multicolumn{4}{c}{AVQA}  \\

% UFA & CFA &
% {\scriptsize $\mathcal{M}_{\mathcal{J}}$} & {\scriptsize $\mathcal{M}_{\mathcal{F}}$} &
% {\scriptsize $\mathcal{M}_{\mathcal{J}}$} & {\scriptsize $\mathcal{M}_{\mathcal{F}}$} &
% % ${\mathcal{J}}$ & ${\mathcal{F}}$ &
% % ${\mathcal{J}}$ & ${\mathcal{F}}$ &
% AQ & VQ & AVQ & Avg. \\

% \midrule
% 1 & 1 & \textbf{80.8} & \textbf{89.9} & 55.4 & 69.9 & 76.1 & 81.2 & 70.9 & 74.5 \\
% 2 & 1 & 80.6 & 89.7 & \textbf{57.1} & 70.3 & 77.7 & 81.8 & 71.2 & 75.2 \\
% 1 & 2 & 80.7 & \textbf{89.9} & 56.4 & \textbf{71.1} & \textbf{78.3} & 81.2 & 71.5 & 75.3 \\

% 2 & 2 & 80.7 & 89.8 & 56.4 & 70.7 & 77.4 & \textbf{83.1} & \textbf{72.3} & \textbf{76.1} \\
% \bottomrule
% \end{tabularx}
% \label{tab:abl_num_expert}
% \end{table}

%% file: figs/fig4_reg.tex
\begin{figure*}[t!]
    \centering
    \includegraphics[width=\textwidth]{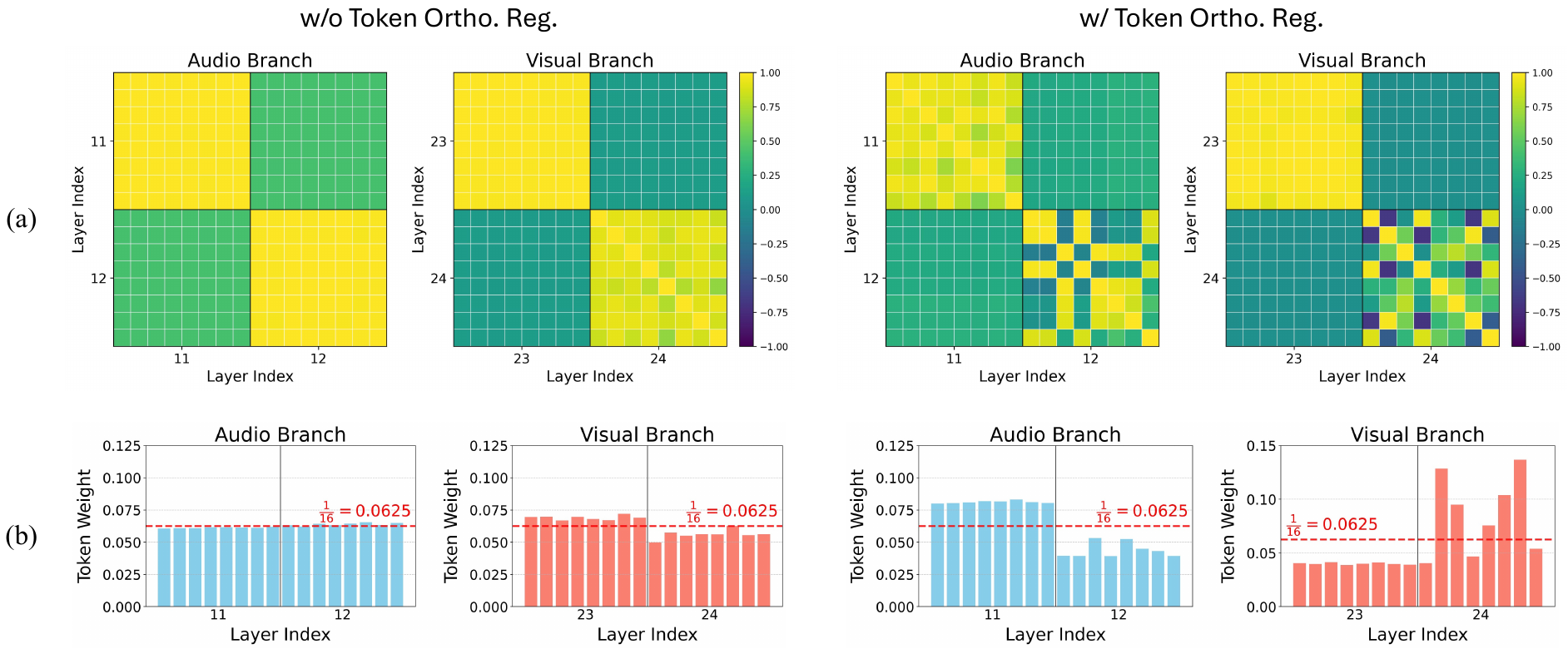}
    \caption{Qualitative results to show the effect of token orthogonality regularization. (a) Cosine similarity map between layer-wise latent tokens, (b) weights for each layer-wise latent token produced by the token fusion module.}
    \label{fig:reg}
    \vspace{-1em}
\end{figure*}

%% file: tabs/abl_fusion.tex
\begin{table}[t!]
\centering
\small
\caption{Ablations on fusion methods for layer-wise latent tokens.}
\begin{tabularx}{\columnwidth}{c|
>{\centering\arraybackslash}X>{\centering\arraybackslash}X>{\centering\arraybackslash}X>{\centering\arraybackslash}X|>{\centering\arraybackslash}X}
% \begin{tabular}{c|cccc|c}

\toprule
% \multicolumn{2}{c|}{\textbf{Num. FDMs}}&
\multicolumn{1}{c|}{Fusion} &
\multicolumn{4}{c|}{AVQA} &
\multicolumn{1}{c}{AVE}  \\

% \textbf{Cross} & \textbf{Uni} &
Method &
AQ & VQ & AVQ & Avg. &
Acc. \\

\midrule
% Avg. pool & - & - & - & - & - & \edit{26g0} & \edit{26g0} & \edit{26g0} & \edit{26g0} \\
% Learnable params & - & - & - & - & - & \edit{26g1} & \edit{26g1} & \edit{26g1} & \edit{26g1} \\
Avg. pool & 76.6 & 82.2 & 71.2 & 75.1        & 82.3 \\
Learnable gates & 75.2 & 80.0 & 70.5 & 73.9          & 80.7 \\
MLP  & \textbf{77.4} & \textbf{83.1}  &  \textbf{72.3}   &  \textbf{76.1}          & \textbf{83.5} \\

\bottomrule
\end{tabularx}
\vspace{-0.5em}
\label{tab:abl_fusion}
\end{table}

% \begin{table*}[t!]
% \centering
% % \renewcommand{\arraystretch}{1.1}
% \caption{Effects of fusion methods for layer-wise latent tokens.}
% \setlength{\tabcolsep}{4.5pt}

% \begin{tabular}{c|c|cc|cc|cccc}
% \toprule
% % \multicolumn{2}{c|}{\textbf{Num. FDMs}}&
% \multicolumn{1}{c|}{Fusion} &
% \multicolumn{1}{c|}{AVE} &
% \multicolumn{2}{c|}{AVS (S4)} &
% \multicolumn{2}{c|}{AVS (MS3)} &
% \multicolumn{4}{c}{AVQA}  \\

% % \textbf{Cross} & \textbf{Uni} &
% Method &
% Acc. &
% $\mathcal{M}_{\mathcal{J}}$ & $\mathcal{M}_{\mathcal{F}}$ &
% $\mathcal{M}_{\mathcal{J}}$ & $\mathcal{M}_{\mathcal{F}}$ &
% AQ & VQ & AVQ & Avg. \\

% \midrule
% % Avg. pool & - & - & - & - & - & \edit{26g0} & \edit{26g0} & \edit{26g0} & \edit{26g0} \\
% % Learnable params & - & - & - & - & - & \edit{26g1} & \edit{26g1} & \edit{26g1} & \edit{26g1} \\
% Avg. pool & - & - & - & - & - & - & - & - & - \\
% Learnable params & - & - & - & - & - & - & - & - & - \\
% MLP & - & - & - & - & - & 77.4  &  \textbf{83.1}  &  \textbf{72.3}   &  \textbf{76.1} \\

% \bottomrule
% \end{tabular}

% \label{tab:abl_fusion}
% \end{table*}

%% file: tabs/abl_reg.tex
\begin{table}[t!]
\centering
\small
\caption{Ablations on the token orthogonality regularization.}
\setlength{\tabcolsep}{2pt}

% \begin{tabular}{r|c|cc|cc|c}
\begin{tabularx}{\columnwidth}{r|
>{\centering\arraybackslash}X|>{\centering\arraybackslash}X>{\centering\arraybackslash}X|>{\centering\arraybackslash}X>{\centering\arraybackslash}X|>{\centering\arraybackslash}X}
\toprule
% \multicolumn{2}{c|}{\textbf{Num. FDMs}}&
%\multicolumn{1}{c|}{Ortho.} 
&
\multicolumn{1}{c|}{AVQA} &
\multicolumn{2}{c|}{AVS-S4} &
\multicolumn{2}{c|}{AVS-MS3} &
\multicolumn{1}{c}{AVE}  \\

% \textbf{Cross} & \textbf{Uni} &
\multicolumn{1}{c|}{Method} &
Acc. &
{\scriptsize $\mathcal{M}_{\mathcal{J}}$} & {\scriptsize $\mathcal{M}_{\mathcal{F}}$}  &
{\scriptsize $\mathcal{M}_{\mathcal{J}}$} & {\scriptsize $\mathcal{M}_{\mathcal{F}}$}  &
Acc. \\

\midrule
w/o $\mathcal{L}_{\text{TOR}}$ & 74.9         & 80.5 & 89.6 & 56.2 & 69.8            & 81.7 \\
w $\mathcal{L}_{\text{TOR}}$  & \textbf{76.1}         & \textbf{80.8} & \textbf{89.9} & \textbf{57.1} & \textbf{70.3}           & \textbf{83.5} \\

\bottomrule
\end{tabularx}

\label{tab:abl_reg}
\vspace{-0.3em}
\end{table}

%% file: sec/5_conclusion.tex
\section{Conclusion}
In this work, we introduce \textbf{MoLT}, a parameter- and memory-efficient framework for audio-visual learning that replaces conventional computationally heavy sequential adaptation with parallel, lightweight layer-wise token distillation and fusion.
MoLT employs a mixture of two types of adapters on frozen backbone layers to distill cross-modal interaction and modality-specific information into layer-wise latent tokens and a token-fusion gating module that aggregates the information from these tokens.
An orthogonality regularization further reduces redundancy among tokens, enabling the fusion module to better capture their relative importance.
Through systematic analysis of the position of adaptation in the pre-trained transformers, we maximize the adaptation effectiveness by only adapting a few last layers of the transformer backbones.
Across a variety of audio-visual downstream tasks, MoLT establishes state-of-the-art performance while substantially reducing trainable parameters and memory usage. 
Ablation studies demonstrate that each component of the proposed MoLT operates effectively, and their combination maximizes adaptation performance.
In the future, we will investigate the effectiveness of the proposed adaptation method in fine-tuning the multimodal large language models.